\documentclass[prl,twocolumn,amsmath,amssymb,showpacs]{revtex4}

\usepackage{graphicx}
\usepackage{dcolumn}
\usepackage{bm}
\usepackage{mathrsfs}
\usepackage{appendix}
\usepackage{bbm}
\usepackage{color}

\def \Cpe {C_{\perp}}
\def \Cpa {C_{\parallel}}
\def \Dfk {\textbf {\emph{D}}_{{\rm FE},k}}
\def \Dak {\textbf {\emph{D}}_{{\rm AFD},k}}

\def \ni {{n_i}}

\def \vS {{\bf S}}
\def \vR {{\bf R}}
\def \ve {{\bf e}}

\def \psc {{\bf P}^{\rm SC}}

\def \pes {{\bf P}^{\rm ES}}

\def \xp {{\bf x}^{\prime }}
\def \yp {{\bf y}^{\prime }}
\def \zp {{\bf z}^{\prime }}
\def \xq {x^{\prime }}
\def \yq {y^{\prime }}
\def \zq {z^{\prime }}

\def \vP {{\bf P}}
\def \vW {{\bf W}}
\def \BF {{\rm BiFeO$_3$} }
\def \BP {{\rm BiFeO$_3$}}

\def \TC {T_c}
\def \TN {T_{\rm N}}

\def \xq {x^{\prime}}
\def \yq {y^{\prime}}
\def \zq {z^{\prime}}
\def \ds {\displaystyle }

\def \ve {{\bf e}}

\def \mB {\mu_{\rm B}}

\begin{document}

{\tiny ``Huge Spin-Driven Polarizations at Room-Temperature in bulk BiFeO$_3$" 
by Jun Hee Lee and Randy S. Fishman. 
This manuscript has been written by UT-Battelle, LLC under Contract No. DE-AC05-00OR22725 
with the U.S. Department of Energy.  The United States Government retains and the publisher, 
by accepting the article for publication, acknowledges that the United States Government retains 
a non-exclusive, paid-up, irrevocable, world-wide license to publish or reproduce the published form 
of this manuscript, or allow others to do so, for United States Government purposes.  The Department of Energy 
will provide public access to these results of federally sponsored research in accordance with the DOE Public Access Plan.}

\title{Huge Spin-Driven Polarizations at Room-Temperature in bulk BiFeO$_3$}
\author{Jun Hee Lee* and Randy S. Fishman}
\affiliation{Materials Science and Technology Division, Oak Ridge National Laboratory, Oak Ridge, Tennessee 37831, USA}
\altaffiliation{e-mail: leej@ornl.gov}

\begin{abstract}

The spin-driven polarizations of type-$I$ multiferroics are veiled 
by the preexisting ferroelectric (FE) polarization. Using first-principles calculations 
combined with a spin model, we uncover two hidden but huge spin-driven polarizations 
in the room-temperature multiferroic \BP. 
One is associated with the global inversion symmetry 
broken by a FE distortion and the other is associated with the local inversion symmetry 
broken by an antiferrodistortive octahedra rotation. Comparison with recent neutron scatterings  
reveals that first polarization reaches $\sim$3.0 $\mu$C/cm$^2$, which is larger than in any other multiferroic material. 
Our exhaustive study paves a way to uncover the various magnetoelectric couplings that generate hidden spin-driven polarizations 
in other type-$I$ multiferroics.

\end{abstract}

\pacs{75.25.-j, 75.30.Ds, 75.50.Ee, 78.30.-j}

\maketitle

Although \BF is endowed with a high ferroelectric (FE) and antiferromagnetic (AF) transition 
temperatures, $\TC \approx 1100$ K \cite{teague70}
and $\TN \approx 640$ K \cite{sosnowska82},
the disparity between $\TC $ and $\TN \ll \TC$ in this type-$I$ multiferroic suggests 
that the magnetoelectric (ME) couplings 
may be quite weak.
Despite enormous effort  \cite{sosnowska82,tokunaga10, park11, lee13,others}, 
a microscopic picture embracing all of the ME coupling mechanisms in bulk \BF is still missing.
By contrast with type-$II$ multiferroics where $\TN = \TC $ and the ME polarizations 
have been well characterized \cite{kimura03},
the large FE polarization, high N\'eel temperature, and long 62 nm period of \BF 
have hindered measurement of its spin-driven ME polarization.
Based on elastic \cite {lee13,arnold09} and inelastic neutron-scattering \cite{inelastic}, 
Raman-scattering \cite{Raman}, and 
THz spectroscopy \cite{spectro} measurements of recently-available single crystals, 
it is now possible to provide detailed information about the intrinsic ME couplings in bulk \BP. 
These results are crucial to control the electrical properties of \BF with a magnetic field and vice versa.

Combining a first-principles approach with a spin-cycloid model, 
we explain the origin of all possible ME couplings and spin-driven (SD) polarizations produced by exchange-striction (ES), 
spin-current (SC), and single-ion anisotropy (SIA).  All polarizations 
are fostered by broken inversion symmetries with two types of lattice distortion 
in $R3c$ bulk \BP: FE and antiferrodistortive (AFD).    
By comparing our results for the spin-driven atomic displacements with elastic neutron-scattering measurements 
\cite{lee13,arnold09,palewicz07}, 
we demonstrate that the ES-polarization (ESP) $\sim$3 $\mu$C/cm$^2$ dominates over other sources of
polarization in the spin cycloid and is larger than any previously reported SD polarization. 

In type-$I$ multiferroics, the absence of an inversion center due to the preexisting FE polarization
fosters the spin-driven polarizatioins.  
Specifically, the change of the scalar product $\vS_i \cdot \vS_j$ at the magnetic transition 
modulates the degree of broken-inversion symmetry and produces corresponding ESPs \cite{FEmagnet}.
While the FE distortion eliminates a global-inversion center, 
the AFD distortion eliminates a local-inversion center. 
Therefore, FE and AF distortions each generate their own ESP. 
 
All possible polarizations are obtained by differentiating the Hamiltonian with respect to an electric field.
For symmetric exchange couplings, ES is dominated by the 
response of the nearest-neighbor interaction $J_1$ from the original Hamiltonian:
\begin{eqnarray}
{\cal H^{\rm EX}}= -\sum_{\langle i,j\rangle }J_1\; \vS_i\cdot \vS_j=-\sum_{\vR_i, \vR_j =\vR_i + \ve_k}J_1^k\; \vS_i\cdot \vS_{j}  ,
\end{eqnarray}
where $k = x$, $y$, or $z$.  Taking the FE polarization along $\zp$=[111], the ESPs are then obtained from
${\textbf {\emph P}^{\rm ES}}=-\vec\nabla_{\vec{E}} {\cal H^{\rm EX}}/N$ with
\begin{eqnarray}
\label{MS1}
{\textbf {\emph P}^{\rm ES}_{\rm FE}} &&= \zp (\zp \! \cdot \! {\textbf {\emph P}^{\rm ES}}) \nonumber \\
    &&= \zp (2C_{\perp}+C_{\parallel}) \zp \cdot \vW_1 \! =\zp \sqrt{3} C_{\rm FE}\, \zp \cdot \vW_1,
\end{eqnarray}
\vspace{-6mm}
\begin{eqnarray}
\label{MS2}
{\textbf {\emph P}^{\rm ES}_{\rm AFD}}&&= \zp \! \times \! {\textbf {\emph P}^{\rm ES}}\nonumber \\
&&  =  (\Cpa \!- \! \Cpe) \zp\! \times \!\vW_2\! = C_{\rm AFD}\, \zp \! \times\! \vW_2 ,
\end{eqnarray}
\vspace{-8mm}
\begin{eqnarray}
 W_{1k}&=&\frac{1}{N} \sum_{\vR_i, \vR_j =\vR_i + \ve_k } \!\vS_i \cdot \vS_j  ,   \\
\vspace{-4mm}
W_{2k}&=& \frac{1}{N} \sum _{\vR_i, \vR_j =\vR_i + \ve_k }  (-1)^\ni \,\vS_i \cdot \vS_j  ,
\end{eqnarray}
where 
$\Cpe=\partial {J_1^{\beta}}/ \partial E_{\alpha}$ ($\beta \neq \alpha$),  
and $\Cpa= \partial J_1^{\alpha} /\partial E_{\alpha}$   
for spin bonds perpendicular and parallel to the electric field, respectively.  

Unlike $W_{1k}$, $W_{2k}$ alternates in sign due to opposite AFD rotations on adjacent hexagonal layers labeled by $n_i$. 
The ESP parallel to $\zp$ with coefficient $C_{\rm FE}=(2\Cpe+\Cpa )/\sqrt{3}$
modulates the FE polarization that already breaks inversion symmetry above $\TN $; 
the ESP perpendicular to $\zp$ has coefficient $C_{\rm AFD}=\Cpe-\Cpa$.  The 
AFD breaks the local inversion between nearest-neighbor spins perpendicular to $\zp $ because each oxygen 
moves along $[0, \overline{1},1]$, $[1,0,\overline{1}]$, and $[\overline{1},1,0]$, perpendicular to $\zp$. 

By ignoring the cycloidal harmonics but including the 
tilt \cite{pyatakov09} $\tau $ produced by $\Dak $,
the spin state propagating along one of the three hexagonal orientaions $\xp$ can be approximated \cite{fishman13b} as
\begin{eqnarray}
\label{syc1}
S_{\xq }(\vR_i)&=& S (-1)^{n_i+1} \cos \tau \sin \big[2 \sqrt{2} \pi \delta \xp\!\cdot\!\vR_i/a\big], \\
\label{syc2}
S_{\yq }(\vR_i)&=& S \sin \tau \sin \big[2\sqrt{2} \pi \delta \xp\!\cdot\!\vR_i/a\big], \\
\label{syc3}
S_{\zq }(\vR_i)&=&S (-1)^{n_i+1} \cos \big[2\sqrt{2} \pi \delta \xp\!\cdot\!\vR_i/a\big],
\end{eqnarray}
where $a$ = 3.96 \AA~is the pseudo-cubic lattice constant and $\delta/(\sqrt{2}a)$ = 62 nm 
is the period of the cycloid so that $\delta \approx 0.0045$.
Recall that \cite{fishman13a} $\sin \tau = S_0/S$ where $M_0=2\mB S_0$ is the weak FM moment of the AF phase 
along $\yp $ above $H_c$.  For a moment \cite{tokunaga10} $M_0=0.03\mB $, $\tau = 0.006$ or 0.34$^\circ $.
Because higher harmonics are neglected, averages taken with the tilted cycloid introduce a very small 
error of order ${C_3}^2 \approx 2.5\times 10^{-5}$.

\begin{figure}
\includegraphics[trim= 1mm 3mm 0mm 0mm, width=8.5cm]{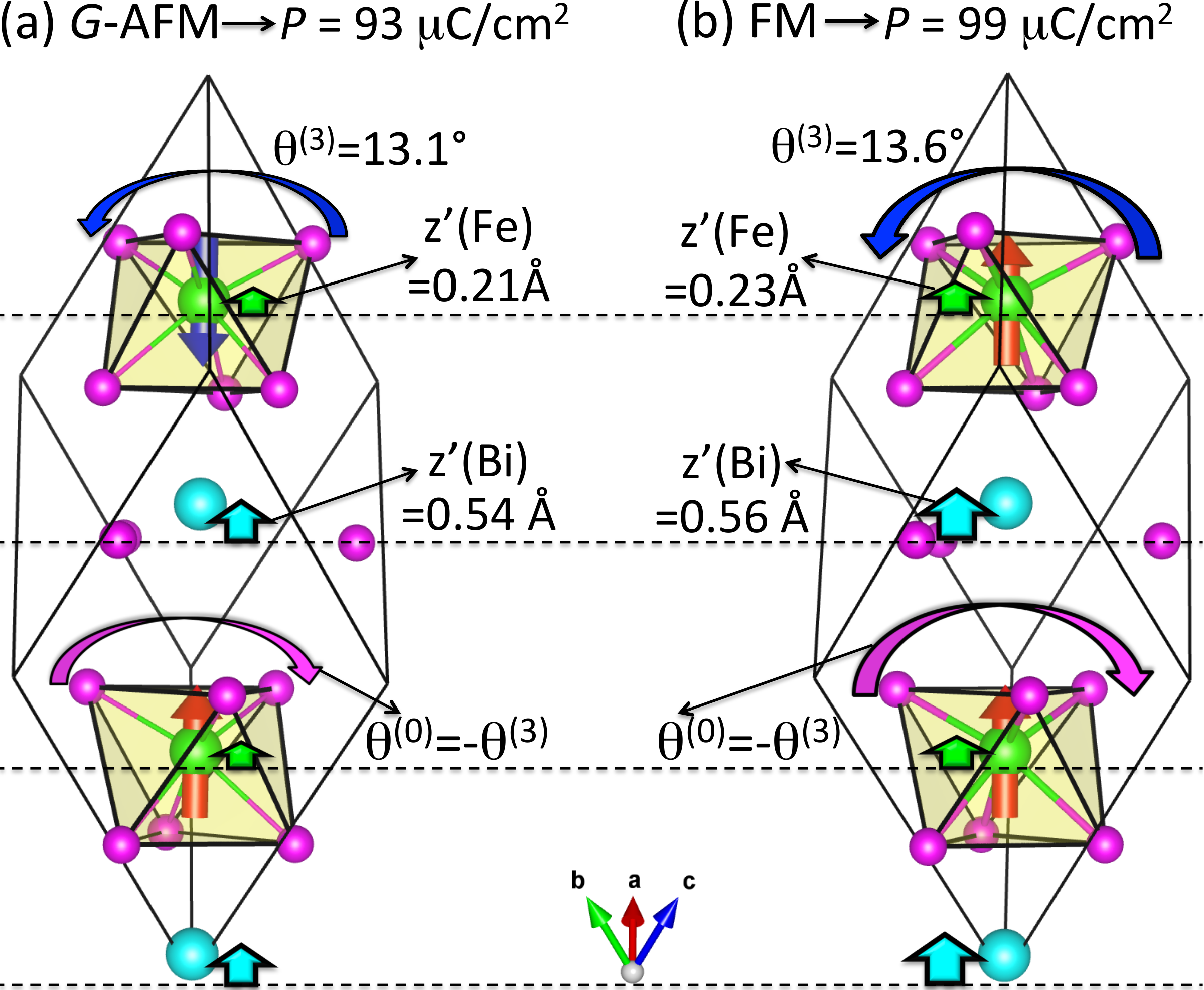}
\caption{ 
First-principles relaxation results for magnetic orderings {\bf (a)} $G$-AFM and {\bf (b)} FM 
to calculate ESPs from FE distortion of Fe ($\zp$(Fe)) and of Bi ($\zp$(Bi))along [111] 
and AFD $R_4^+$ rotations ($\theta$) perpendicular to [111].
(0) and (3) are indices of hexagonal layers ($x+y+z=n$).}
\label{MS}
\end{figure}

To calculate the electric-field induced lattice distortion 
and the associated change in $J_1$, we employ a method \cite{iniguez08} that was successfully 
applied to other multiferroic oxides.  The resulting ESP coefficients are given in Tab.~\ref{J1}. 
Using the result $C_{\rm FE} =215$ nC/cm$^2$ and Eq.(\ref{MS1}) for $\pes_{\rm FE}$,
we find that the ESP for the simple twisted cycloid in Eqs.(\ref{syc1} - \ref{syc3}) is 
\begin{equation}
\label{ms1}
\langle \pes_{\rm FE} \rangle = -C_{\rm FE} S^2 \zp \cos^2 \tau = -1.3 \; \mu{\rm C/cm}^2 \zp . 
\end{equation}
Because a harmonic approximation \cite{iniguez08} was used to generate the possible polar distortions 
induced by the electric field, this ESP was only evaluated to quadratic order in the lattice distortions driven by the spin ordering.

However, the ESP may be large enough to induce atomic displacements and lattice distortions beyond the harmonic limit. 
As shown in Fig.~\ref{MS}, one can calculate the ESP more accurately including anharmonic effects and spin-lattice couplings 
by fully relaxing the atoms and the lattice
for different magnetic orderings ($G$-AFM and FM) with
\begin{eqnarray}
\Delta P_{\rm FE}^{\rm ES}&&=P_{\rm FM}-P_{\rm AFM}=2C_{\rm FE}S^2  \hspace{50pt} \nonumber \\
&& =\frac{1}{V}\!\!\sum_{i={\rm Fe,Bi}} \!\!\{Z^*_{i, \rm FM} u_{i, \rm FM}\!-\!Z^*_{i, \rm AFM} u_{i, \rm AFM}\}  \nonumber  \\ 
\vspace{-1mm}
&& =6.0 \; \mu{\rm C/cm}^2, 
\label{largeC}
\end{eqnarray}
so that $C_{\rm FE}=480\;{\rm nC/cm}^2$ and   
\begin{equation}
\langle \pes_{\rm FE} \rangle = -C_{\rm FE} S^2 \zp \cos^2 \tau = -3.0 \; \mu{\rm C/cm}^2 \zp, 
\label{MS11}
\end{equation}
where $Z^*_i$, $u_i$, and $V$ represent Born effective charge, atomic position and volume, respectively. 
The change of spin ordering from $G$-AFM to FM shifts the Fe and Bi atoms by 0.020 \AA ~and 0.019 \AA , respectively as shown in Fig.~\ref{MS}. 
While the Bi effective charge hardly changes (Z$_{\rm AFM}^*$=4.82e, Z$_{\rm FM}^*$=4.83e), 
the Fe effective charge changes significantly (Z$_{\rm AFM}^*$=3.91e, Z$_{\rm FM}^*$=4.11e) 
due to the spin-induced hybridization between Fe and oxygen. 
Consequently, C$_{\rm FE}$=480 nC/cm$^2$ is a factor of two larger than
the harmonic value in Tab.~\ref{J1}.  This result will later be compared with neutron-scattering measurements.

\begin{table}
\caption{Calculated (LSDA+$U$) ESP components 
perpendicular $\Cpe$ or parallel $\Cpa $ to the electric field and associated
FE ($C_{\rm FE}$) and AFD ($C_{\rm AFD}$) spin-driven polarization. 
Values in the parenthesis are directly obtained from the polarization difference of the relaxed structures 
with $G$-AFM or FM ordering as shown in Eq.~\ref{largeC}, \ref{MS22} and Fig.~\ref{MS}.}
\begin{ruledtabular}
\begin{tabular}{ccccccc}
nC/cm$^2$ & $\Cpe$ &$\Cpa$ & $C_{\rm FE}$ &  $C_{\rm AFD}$ \\
 \hline                                              
LSDA+$U$  &  186    &  0.769  &   215 (480)   &  -185 (-108)    \\
\end{tabular}
\end{ruledtabular}
\label{J1}
\end{table}

\begin{figure*}
\includegraphics[trim= 1mm 2mm 0mm 0mm, width=17.6cm]{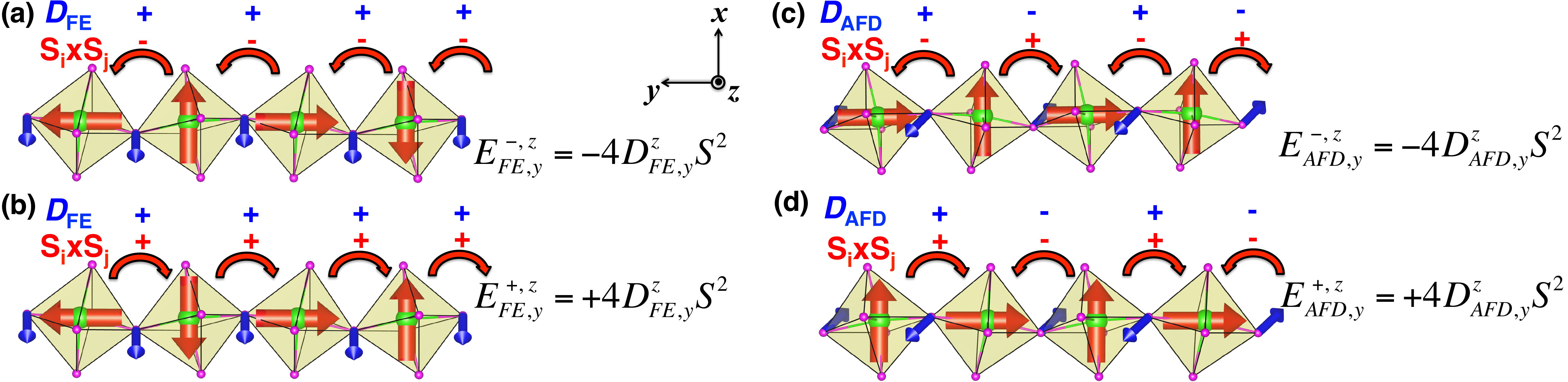}
\caption{Separation of FE ($\Dfk $) and AFD ($\Dak $) DM interactions in \BP.
Spin directions (red arrows) and DM vectors (blue arrows) are indicated.
{\bf (a)} and {\bf (b)} have clockwise and counter-clockwise rotation of spins and use an 80 atom unit-cell 
(four spins in the unit-cell along each direction). 
 {\bf (c)} and {\bf (d)} have zigzag-type spin ordering.
$+/-$ above red arrows denote signs of DM interaction and of spin-current between 
two nearest neighbor spins.}
\label{fig_D}
\end{figure*}

The Supplement shows that Eq.(\ref{MS2}) describes the ESP perpendicular to the FE-polarization direction $\zp$
due to the inversion-symmetry breaking from AFD rotations.
Hence, the second ESP reflects the change of AFD and the associated ${\emph local}$ polarization 
driven by spins perpendicular to $\zp$.  
For a simple tilted cycloid, $\langle W_{2k} \rangle = \langle P^{\rm ES}_{\rm AFD} \rangle =0$  
because the AFD distortions do not globally break inversion symmetry. 

As shown in Fig.~\ref{MS}, first-principles calculations can capture $P^{\rm ES}_{\rm AFD}$
by evaluating the AFD-induced oxygen displacements perpendicular to $\zp$ with the change of spin ordering. 
The increase of the AFD rotation ($\Delta \theta=0.54^\circ$) from $G$-AFM ($\theta = 13.08^\circ$) 
to FM ($\theta = 13.62^\circ$) corresponds to an increase of the oxygen displacement (0.015\,\AA) 
along $[0,-1,1]$, $[1,0,-1]$, and $[-1,1,0]$ perpendicular to $\zp$.  Therefore,
\begin{eqnarray}
\label{MS22}
\Delta P_{\rm AFD}^{\rm ES}&=&P_{\rm FM}-P_{\rm AFM}=2C_{AFD} S^2 \nonumber \\
         &=&\frac{-3.30e}{V}0.015\AA = -1340 \; {\rm nC/cm}^2 , 
\end{eqnarray}
so that          
$C_{\rm AFD}=-108 \; {\rm nC/cm}^2$.

Interestingly, the two ESPs C$_{\rm FE}$ and C$_{\rm AFD}$ are coupled.
While the ESP components $\Cpe $ and  $\Cpa$ cooperatively 
increase ESP along $\zp$ under the inversion symmetry broken by FE, they produce opposite contributions 
to the AFD-induced ESP perpendicular to $\zp$.
$\Cpe$ is largely positive due to the reduction of the Fe-O-Fe bond angle driven by the FE distortion, which favors 
FM from the Goodenough-Kanamori (GK) rules \cite{goodenough}; 
$\Cpa$ is almost zero because the bond contraction between Fe-O-Fe does not significantly alter 
the spin-density environment around the $d^5$ electrons of Fe. The large difference between $\Cpe$ and $\Cpa$ induces 
the large ESP induced by AFD rotations.  

The global and local inversion symmetry breaking by FE and AFD distortions produce the DM interactions 
$\Dfk $ and $\Dak $.  
Due to their distinct translational characters, they can be separated using the procedure sketched in Fig.~\ref{fig_D}.
Since the FE distortion is globally uniform, 
its associated $\Dfk $ is uniform too. 
Because the AFD rotation alternates between hexagonal layers, 
the associated DM vector $\Dak $ also alternates as shown by the blue arrows in Fig.~\ref{fig_D}. 
As shown in Figs.~\ref{fig_D}(a) and (b), a cycloid consisting of four spins along $\ve_k$ 
generates a translation-invariant spin current $\vS_i$ $\times$ $\vS_j$.
The uniform component $\Dfk $ is extracted from
\vspace{-1mm}
\begin{eqnarray}
E^{\pm,\gamma}_{{\rm FE},k} = E_0 \pm 4D^{\gamma}_{{\rm FE},k}S^2-\frac{4}{3}KS^2,
\end{eqnarray}
\vspace{-7mm}
\begin{eqnarray}
D^{\gamma}_{{\rm FE},k} = \frac{1}{8S^2}(E_{{\rm FE},k}^{+,\gamma}-E_{{\rm FE},k}^{-,\gamma}),
\end{eqnarray}
where $\pm$ refer to counterclockwise (+) and clockwise ($-$) rotations, respectively.   
The translation-odd $\Dak $ does not appear in this expression.

Using a zig-zag type spin arrangement that generates a spin current $\vS_i$ $\times$ $\vS_j$ 
with alternating sign, the translation-odd $\Dak $ is extracted from
\vspace{-1mm}
\begin{eqnarray}
E^{\pm,\gamma}_{{\rm AFD},k} = E_0 \pm 4D^{\gamma}_{{\rm AFD},k}S^2-\frac{4}{3}KS^2,
\end{eqnarray}
\vspace{-7mm}
\begin{eqnarray}
D^{\gamma}_{{\rm AFD},k} = \frac{1}{8S^2}(E_{{\rm AFD},k}^{+,\gamma}-E_{{\rm AFD},k}^{-,\gamma}),
\end{eqnarray}
which does not contain the translation-even $\Dfk $.

As for the ESP, the SC polarization (SCP) $\psc = \psc_{\rm FE} + \psc_{\rm AFD}$ splits into terms 
produced by the inversion-symmetry breaking of FE and AFD distortions, respectively. 
The first SCP is caused by the response of the FE distortion to an electric field:
\begin{eqnarray}
&&P^{SC}_{\rm FE,\gamma}= -\frac{\partial {\cal H_{\rm FE}^{\rm SC}}}{\partial E_{\gamma}}   \nonumber \\  
 &&= -\frac{1}{N}\ds\sum_{\ve_k} \frac{\partial{\textbf {\emph{D}}_{{\rm FE},k}}}{\partial E_{\gamma}} 
\cdot \bigl(\vS_i\times \vS_{i+\ve_k}\bigr) . 
\end{eqnarray}
The $E$-field derivatives of the DM, ${\bf f}^{k\beta }= \partial  {\textbf {\emph{D}}_{{\rm FE},k}}/\partial E_{\beta }$
are presented in Tab.~\ref{all} and the Supplement

The second SCP arises from the AFD rotation.  Its sign alternates due to the alternating AFD rotation direction along [111]:
\begin{eqnarray}
&&P^{SC}_{\rm AFD,\gamma} =-\frac{\partial {\cal H_{\rm AFD}^{\rm SC}}}{\partial E_{\gamma}}  \nonumber \\
 &&=-\frac{1}{N}\ds\sum_{\ve_k} (-1)^{n_i} \frac{\partial{\textbf{\emph{D}}_{{\rm AFD},k}}}{\partial E_{\gamma}} 
 \cdot  \bigl(\vS_i \times \vS_{i+\ve_k}\!\bigr) . 
\end{eqnarray}
The SCP components ${\bf a}^{k\beta }=\partial {\textbf{\emph{D}}_{{\rm AFD},k}}/\partial E_{\beta }$ 
are evaluated in the Supplement and presented in Tab.~\ref{all}. 

\begin{figure}
\includegraphics[trim= 0mm 7mm 0mm 0mm, width=6.9cm]{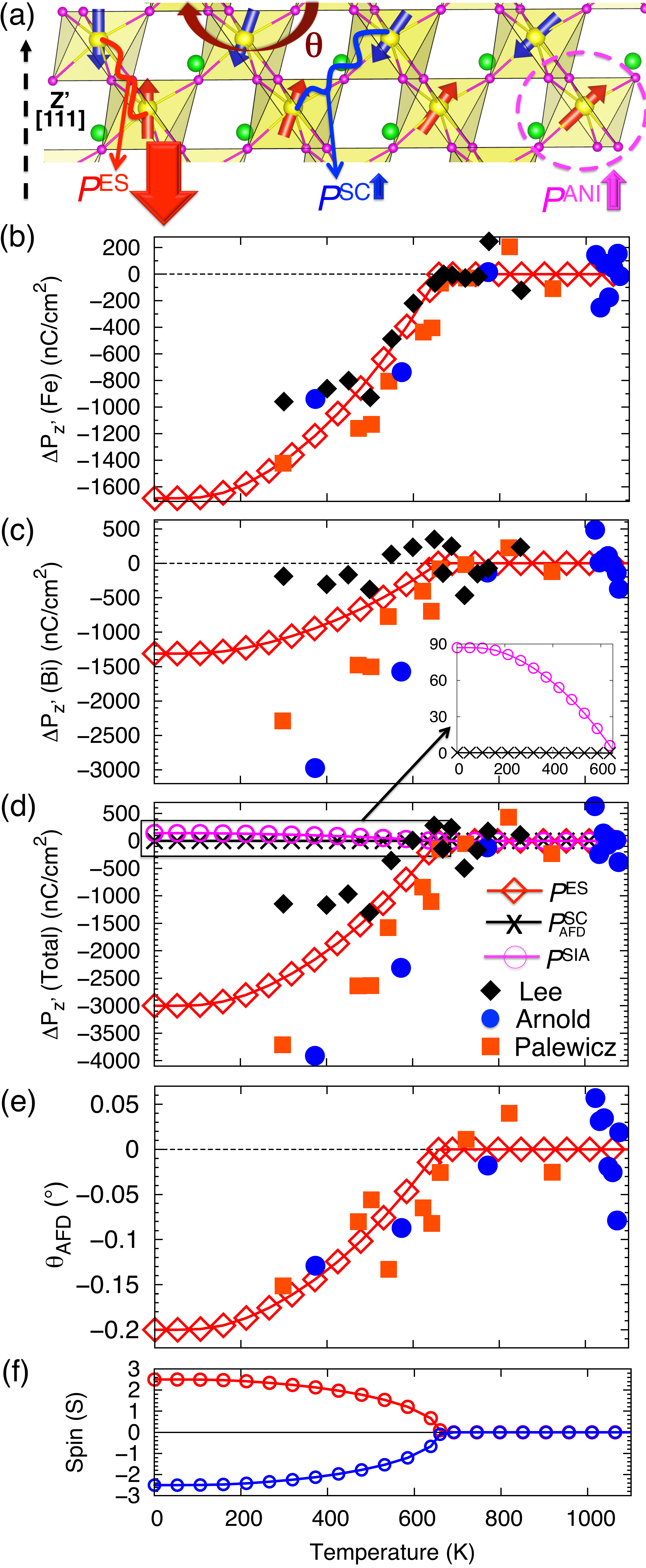}
\caption{
{\bf (a)} Spin-driven polarizations along $\zp$ and AFD rotations perpendicular to $\zp$. 
{\bf (b)} Calculated Fe polarizations along $\zp$ induced by ES ($P^{\rm ES}$) 
compared with measurements by Lee {\it et al.} [\onlinecite{lee13}], Arnold {\it et al.} [\onlinecite{arnold09}], 
         and Palewicz {\it et al.} [\onlinecite{palewicz07}]. 
{\bf (c)} Bi polarizations along $\zp$ induced by ES.  
{\bf (d)} Total SD polarization $\Delta P_{\zq }$(Fe)+$\Delta P_{\zq }$(Bi). 
Inset is the SC polarization calculated from Eqs.(\ref{DFE}-\ref{DAFD}) and Ref.[\ref{SIA}]. 
{\bf (e)} Change of AFD-rotation angle induced by ES. 
{\bf (f)} AFM spin ordering versus temperature calculated by mean-field theory.} 
\label{ES}
\end{figure}

SIA is the last possible source for the ME polarization.  
Starting with the SIA enenrgy ${\cal H^{\rm SIA}}= -K\sum_i(\vS_i\cdot\zp )^2$, 
the SD-polarization has $\zp$ component
\begin{eqnarray} 
\zp \! \cdot \! {\textbf {\emph P}^{\rm SIA}}
    =\frac{\sqrt3 \xi}{N}  \sum_i (\vS_i \cdot \zp )^2, \;\; \xi=\frac{\partial K}{\partial E_{\alpha}} . 
\end{eqnarray}
The elongation $\xi$ of $K $ is given in Tab.~\ref{all}. 

\begin{table}
\caption{Calculated (LSDA+$U$) SCPs and SIA polarization (SIAP)
derived from DM and $K$ interactions.
Subscripts denote whether the spin-bond direction is parallel or perpendicular 
to the electric field. $f_{z}^{xy}$, $f_{y}^{xy}$ are not shown due to low convergence.}
\begin{ruledtabular}
\begin{tabular}{cccccccc}
          &           \multicolumn{6}{c}{SCP from $D$}       & SIAP  \\
nC/cm$^2$ & $a_{x}^{xx}$ & $a_{y}^{xx}$ & $a_{y}^{xy}$ & $a_{x}^{xy}$ & $a_{z}^{xy}$ & $f_{y}^{xy}$ & $\xi$\\
 \hline                                              
LSDA+$U$ & 6.6 & -21 & 16 & 1.2 & -13 & -4.9 & 16 \\
\end{tabular}
\end{ruledtabular}
\label{all}
\end{table}

For the simple tilted cycloid of Eqs.(\ref{syc1}-\ref{syc3}), 
\begin{equation}
\langle \vP^{{\rm SIA}}\rangle = \frac{\sqrt3}{2N}\xi S^2 \; \zp  ,
\label{SIA}
\end{equation}
\vspace{-7mm}
\begin{eqnarray}
\langle \psc_{\rm FE}  \rangle =-2\pi\sqrt6 S^2 \; \delta \cos \tau \hspace{99pt} \nonumber  \\
\vspace{-15mm}
\biggl\{(f_{y}^{xx}\!-\!f_{y}^{xy})\;\yp
+\sqrt2(f_{z}^{xy}\!-\!f_{y}^{xy}\!-\!f_{y}^{xx})\;\zp\biggr\},
\label{DFE}
\end{eqnarray}
\vspace{-6mm}
\begin{eqnarray}
\langle \psc_{\rm AFD} \rangle\!\! =\!\! -S^2 \! \sin \! 2\tau  \biggl\{\!\frac{a_{x}^{xx}}{2} \! \!
    + a_{y}^{xx}\! \! +a_{x}^{xy} \! \!  
    +a_{y}^{xy} \! \! +a_{z}^{xy}\!\! \biggr\}  \; \zp .
\label{DAFD}
\end{eqnarray}
Symmetry relations for $f^{\alpha\beta}_{\gamma}$ and $a^{\alpha\beta}_{\gamma}$ are given in the Supplement.
Because $\delta,\tau \ll 1$, the projected polarization along $\zp$ 
is $\zp \cdot \langle {\textbf {\emph{P}}^{\rm SC}}+{\textbf {\emph{P}}^{\rm SIA}}\rangle 
 \approx \zp \cdot \langle {\textbf {\emph{P}}^{\rm SIA}}\rangle \approx 87$ nC/cm$^2$ which is larger than the 
experimental value 40 nC/cm$^2$ \cite{park11,tokunaga10}
obtained from the jump in polarization below the critical field $H_c \approx 20$ T.
Intriguingly, $\Dfk $ produces an additional SCP along $\yp$, which may 
explain why the SC also generates a polarization perpendicular to $\zp$ \cite{park11}.
Since the SCP and SIAP are still much smaller than the ESP,
the dominant polarization at the magnetic transition is driven by ES. 

Figure~\ref{ES} shows all the ME polarizations driven by the AFM spin-ordering around $\TN $ and compares those results
to elastic neutron-scattering measurements \cite{lee13, arnold09, palewicz07}. 
Although the neutron-scattering data is rather spread, all three papers 
indicate that both the polarization and AFD rotation angle are reduced 
by the huge ES around $\TN $. 
As explained in the Supplement, we convert the preliminary neutron-scattering data 
to the change of the spin-driven polarization at $\TN$ using Ginzburg-Landau free energies.
With spin ordering (Fig.~\ref{ES}(d)), Fe and Bi move $-0.010 \AA $ and $-0.0095 \AA $, respectively, and 
induce polarizations $\Delta P$(Fe)= $-1.7~\mu$C/cm$^2$ and $\Delta P$(Bi) = $-1.3~\mu$C/cm$^2$. 
The net induced polarizaton $P_{\rm tot}$ = $\Delta P$(Fe)+$\Delta P$(Bi) = $-3.0~\mu$C/cm$^2$ is in excellent agreement with 
neutron-scattering measurements. 

Intriguingly, both Bi and Fe shift with the spin ordering. 
Fe moves antiparallel to reduce the FE polarizations because AFM ordering favors $180^{\circ}$ displacements by the
GK rules \cite{goodenough}. The reduction in the Fe polarizations simultaneously reduces the Bi polarization.
Consequently, the magnitude of the net polarization is greater than any previously-reported spin-driven polarization
(0.29 $\mu$C/cm$^2$ in CaMn$_7$O$_{12}$ \cite{johnson12} and 0.36 $\mu$C/cm$^2$ in GdMn$_2$O$_5$ \cite{2lee13}). 

We have also discovered another huge but hidden ES due to AFD rotations that are strongly coupled to the FE-driven ES. 
Obviously, this contribution cannot be easily measured 
because AFD rotations do not break global inversion symmetry and do not produce a net macroscopic polarization.
However, AFD rotations do break local inversion symmetry and their associated atomic displacements appear
in the neutron-scattering data in Fig~\ref{ES}(e). 

Based on the good agreement between our predictions and neutron-scattering results, 
we conclude that the AFD rotation angle is suppressed by ES.
Both the polarization and AFD reduction can be understood in terms of the GK rules \cite{goodenough}:
AFM ordering decreases bond angles, which reduces the FE polarizaton and AFD rotations.  
Due to recent advancement in local polarization measurements \cite{probe}, 
it may soon be possible to directly image the spin-driven structural modification of the AFD.


Although some calculations predict polarizations $\sim 6~\mu$C/cm$^2$ \cite{ivan06,picozzi07} 
for orthorhombic perovskites such as HoMnO$_3$, 
the largest measured spin-driven polarization prior to this work was found in pressurized TbMnO$_3$, where $P= 1.0~\mu$C/cm$^2$
can rise to $1.8~\mu$C/cm$^2$ with 5.2 GPa \cite{aoyama14} at 5 K. 
We have checked the possible spin-driven polarization 
in another type-$I$ multiferroic, BiCoO$_3$, from neutron scattering \cite{BiCoO3}.
But its polarization appears to be smaller than that of \BF by one order of magnitude. 
Hence, the type-$I$ multiferroic \BF unexpectedly exhibits the largest ever spin-driven polarization ($\sim 3~\mu$C/cm$^2$) at room temperature. 
There are three reasons for this huge spin-driven polarization.
First, the exchange interactions and their response to external perturbations such as electric field or temperature 
are larger than for the DM interactions, as shown in Tab.~\ref{J1} and Tab.~\ref{all}. 
Second, even if the ES coefficients were the same for \BF and the above manganites, the enhanced spin correlation function 
($\mid$$\langle\vS_i\cdot \vS_j\rangle$$\mid$=$S^2$=6.25 in \BF, and 2.25 in the manganites) would strongly enhance the ESP in \BP. 
Third, in contrast to the $E$-type ordering in the manganites, the almost antiparallel alignment of neighboring spins in \BF also enhances the ESP.  

The greatest advantages of \BF are its large FE polarization, high $\TC $, and high $\TN $ above room temperature. These 
advantages have unfortunately hampered precise characterizations 
of the ME polarizations around $\TN$. Leakage currents at high temperatures and the preexisting large FE polarization 
have hidden the spin-driven ME polarizations at $\TN$. 
Fortunately, intrinsic measurments such as neutron-scattering, Raman spectroscopy and directional dichroism have recently
begun uncovering the hidden ME couplings of \BP. 
So in addition to having the largest known FE polarization, \BF may also have spin-driven polarizations 
much larger than in any other known material. Our systematic approach will greatly aid further exploration of hidden but
possibly large spin-driven polarizations and their ME origins in other type-$I$ multiferroics. 

We appreciate S. Lee, J.-G. Park, and S. Okamoto for valuable discussions. 
Research sponsored by the U.S. Department of Energy, Office of Science, 
Office of Basic Energy Sciences, Materials Sciences and Engineering Division and by the Scientific
User Facilities Division, Office of Basic Energy Sciences, US
Department of Energy.

\end{document}